\documentclass[twocolumn,showpacs,preprintnumbers,amsmath,amssymb]{revtex4}

\usepackage{graphicx}
\usepackage{dcolumn}
\usepackage{bm}

\begin{document}

\preprint{APS/123-QED}

\title{Influence of phonons on exciton-photon interaction and photon statistics of a quantum dot}

\author{M. Bagheri Harouni, R. Roknizadeh and M. H. Naderi}
\affiliation{$^1$Quantum Optics Group, Department of physics,
University of Isfahan, Isfahan, Iran}

\date{\today}

\begin{abstract}
In this paper, we investigate, phonon effects on the optical
properties of a spherical quantum dot. For this purpose, we
consider the interaction of a spherical quantum dot with classical
and quantum fields while the exciton of quantum dot interacts with
a solid state reservoir. We show that phonons strongly affect the
Rabi oscillations and optical coherence on first picoseconds of dynamics.
We consider the quantum statistics of emitted photons by quantum
dot and we show that these photons are anti-bunched and obey the
sub-Poissonian statistics. In addition, we examine the effects of
detuning and interaction of quantum dot with the cavity mode on
optical coherence of energy levels. The effects of detuning and
interaction of quantum dot with cavity mode on optical coherence of energy
levels are compared to the effects of its
interaction with classical pulse.
\end{abstract}

\maketitle
\section{Introduction}
The fundamental system in cavity quantum electrodynamics
(cavity-QED) is a two level atom interacting with a single-cavity
mode \cite{qo1}-\cite{qo2}. Recent developments in semiconductor
nano-technology have shown that excitons in quantum dots (QDs)
constitute an alternative two-level system for cavity-QED
application \cite{gerard}. There are many similarities between the
excitons in QDs and atomic systems, such as the discrete level
structures which is subsequent of three-dimensional confinement of
electrons. On the other hand, there are also important
differences, for example coupling to phonons, carrier-carrier
interaction and surface fluctuation. Coupling of electrons to
phonons plays a major role in QDs. The coupling of phonons to the
QD provides a basic dephasing mechanism and thus marks a lower
limit for the decoherence \cite{mulj}-\cite{knorr}. In
self-assembled QDs it is indeed the elastic phonon scattering
(pure dephasing) which dominates the loss of coherence on a
picosecond time scale at temperatures below $100$K \cite{borri}.
The effects of electron-phonon interactions on strong
exciton-photon coupling in cavity-QED has been considered
\cite{wilson}. It has been shown \cite{vagov2} that the
phonon-induced damping of Rabi oscillations in a QD is a
non-monotonic function of the laser-field intensity that is
increasing at low fields and decreasing at high fields.\\ \indent
QDs are also promising candidates for efficient, deterministic
single photon sources \cite{shield}-\cite{santor}. Then the QDs
are important sources of non-classical light. For this kind of
application an understanding of the coherence properties of its
optical transitions is of great importance. Therefore, there are
two processes in optical manipulation with semiconductor QDs:
coherent control of the QD exciton state \cite{axt} and
measurement of quantum statistics of emitted light with QD
\cite{muller}. A theoretical investigation of exciton dynamics and
the possibility of generation of non-classical light has been
considered without taking into account the phonon effects \cite{perea}.\\
\indent In this paper, we investigate the effects of electron-phonon
interactions on optical coherence and quantum statistics of light
emitted by a pulse driven QD interacting with a cavity mode. The
photon statistics from a driven QD under the influence of the
phonon environment has been considered recently \cite{zahir}. On the other hand, 
influence of phonons on incoherent photon emission of a QD in the presence of 
pulse excitation had been considered \cite{ahn}. We
use the most widely studied model for phonon effects in QDs which
accounts two electronic levels coupled to a laser pulse and to
non-interacting phonons \cite{machni}. As mentioned, phonon
interaction provides a dephasing mechanism for optically induced
coherence on a time scale (a few picosecond) much shorter than for
radiative interaction and recombination \cite{borri1}. Due to the
different correlation time for a phonon reservoir (few picosecond)
and for a radiative reservoir (several ten nanosecond) we restrict
our attention to the time scales which dephasing effects due to
the phonon system play an important role (with the radiative reservoir 
we mean a reservoir for photon system. The mentioned time scale relates to 
the decay time of cavity photons). Then we do not consider
any damping effect on cavity mode and spontaneous emission. In our
consideration the only damping effect is related to phonons. \\
\indent The paper is organized as follows: In section II we
describe the model Hamiltonian and master equation that allows to
calculate the evolution of populations and coherence of the energy
levels. In section III we present the exciton dynamics and its
coherence while driven with a laser pulse. The photon statistics
and exciton dynamics of pulse driven QD interacting with a cavity
mode is presented and discussed in section IV. Section V is
devoted to a summery and conclusion.
\section{Theoretical model}
We consider a single QD inside a semiconductor microcavity that is
pumped with a laser pulse and interacts with a cavity mode. It is
assumed that the system is initially prepared in its ground state.
We consider a solid-state reservoir for the exciton population and
we focus on time scales which phonon effects are important. We
neglect other sources of damping in the system. We model the QD by
a two-level system with ground state $|g\rangle$ (the
semiconductor ground-state) and first excited state $|e\rangle$ (a
single exciton), separated by an energy $\hbar\omega_{ex}$. The
phonon environment is modelled by a bath of harmonic oscillators
of frequencies $\omega_k$, with the wavevector $k$. The
Hamiltonian of the total system in the rotating wave approximation is
written as
\begin{eqnarray}\label{}
    \hat H&=&\hbar\omega_{ex}\hat\sigma_{ee}+\hbar\omega_c\hat a^{\dag}\hat a+\sum_k\hbar\omega_k\hat b^{\dag}_k\hat b_k\nonumber \\ &+&\hbar
    g(\hat\sigma_{eg}\hat a+\hat a^{\dag}\hat\sigma_{ge})+\hbar
    f(t)(\hat\sigma_{eg}+\hat\sigma_{ge})\nonumber \\&+&\hat\sigma_{ee}\sum_k\lambda_k(\hat b_k+\hat
    b^\dag_k),
\end{eqnarray}
where $\hat\sigma_{ij}=|i\rangle\langle j|$, $\hat a\;(\hat
a^\dag)$ and $\hat b_k\;(\hat b^\dag_k)$ are the annihilation
(creation) operators for cavity mode, and $k$th phonon mode,
respectively. The parameter $g$ is the coupling constant of the
exciton and cavity mode, and $f(t)$ is a real envelope function of
the driving pulse. The last term in the Hamiltonian describes the
exciton-phonon interaction. In this term, $\lambda_k$ is the
corresponding coupling constant. The coupling of the confined
exciton to the acoustic phonons by means of the deformation potential
tends to dominant the dephasing dynamics, over the piezoelectric
interaction or coupling to optical phonons \cite{krumm}. In this
case, the coupling constant is given by
$\lambda_k=kD(k)\sqrt{2n\omega_kV}$ \cite{mahan} where $n$ is the
sample density and $V$ is the unit cell volume. $D(k)$ is the form
factor of the confined electron and hole in the ground state of
the QD. The Hamiltonian in the interaction picture can be written
as
\begin{equation}\label{}
    \hat H_{int}=\hat H_0+\hat H_R,
\end{equation}
where we decompose the coherent-field part and environment part as follow
\begin{eqnarray}\label{}
    \hat H_0&=&\hbar g(\hat\sigma_{eg}\hat ae^{i\Delta t}+\hat a^{\dag}\hat\sigma_{ge}e^{-i\Delta
    t})\nonumber \\&+&\hbar
    f(t)(\hat\sigma_{eg}e^{i\omega_{ex}t}+\hat\sigma_{ge}e^{-i\omega_{ex}t}),\nonumber
    \\ \hat H_R&=&\hat\sigma_{ee}\sum_k\lambda_k(\hat b_ke^{-i\omega_kt}+\hat
    b^{\dag}e^{i\omega_kt}).
\end{eqnarray}
In this equation $\Delta=\omega_{ex}-\omega_c$ is detuning between the exciton excitation energy in the QD and cavity field energy.\\
Now we consider the Liouville equation of density matrix in the
interaction picture
\begin{equation}\label{}
    \frac{d\hat\rho_t}{dt}=\frac{i}{\hbar}[\hat\rho_t,\hat
    H_{int}].
\end{equation}
We define the reduced density matrix $\hat\rho$ for the
exciton-photon system by tracing out the phonon degrees of freedom
in the total density matrix, $\hat \rho=Tr_{ph}(\hat \rho_t)$. Now
we consider the master equation in the Born approximation
\cite{qo1}-\cite{qo2} in the case of the phonon interaction while we
consider the gain and pump parts exactly. Phonons are one of the slowest
 process and this kind of reservoir has a correlation time of the order of 
a few picosecond \cite{krumm} and this reservoir is naturally non-Markovian. 
To consider non-Markovian dynamics we have used time convolutionless projection 
operator method \cite{bruer}, up to second order of expansion. We assume an
uncorrelated state for initial state of the exciton-photon system
and phonon reservoir. At the initial time $t=0$ the phonon system
is assumed to be in a thermal equilibrium at temperature $T$. Then
the density operator of the exciton-photon system satisfies the
following dynamical equation
\begin{eqnarray}\label{eq1}
    \dot{\rho}(t)&=&\frac{i}{\hbar}[\rho(t),\hat H_0]-\int_0^t
    ([\hat\sigma_{ee},\hat\sigma_{ee}\rho(t)]K(t-t')\nonumber \\ &-&[\hat\sigma_{ee},\rho(t)\hat\sigma_{ee}]K^\ast(t-t'))dt'.
\end{eqnarray}
The first term describes the coherent evolution of the density matrix $\rho$ under the action of the Hamiltonian $\hat H_0$ of the dot-cavity-pulse system. The kernel $K$ which is the correlation function of the
environment is written as
\begin{equation}\label{}
    K(t)=\frac{1}{\hbar^2}\int_0^\infty d\omega
    j(\omega)\left[coth(\frac{\hbar\omega}{2k_BT})cos(\omega t)-isin(\omega
    t)\right],
\end{equation}
with Boltzmann constant $k_B$. $j(\omega)$ is the spectral density
of the phonons which completely describes the interaction of
exciton and phonons \cite{weiss}. Here, we introduce the following
spectral density
\begin{equation}\label{}
    j(\omega)=\sum_k\lambda_k^2\delta(\omega-\omega_k).
\end{equation}
The density matrix dynamics is obtained under the Born-Markov
approximation for exciton-phonon interaction and the strong
exciton-photon interaction and pump effects are described exactly.
We can extract exciton dynamics and photon statistics from this
equation.
\section{Exciton dynamics under a driving pulse}
\begin{figure*}
\begin{center}
\includegraphics[angle=0,width=.5\textwidth]{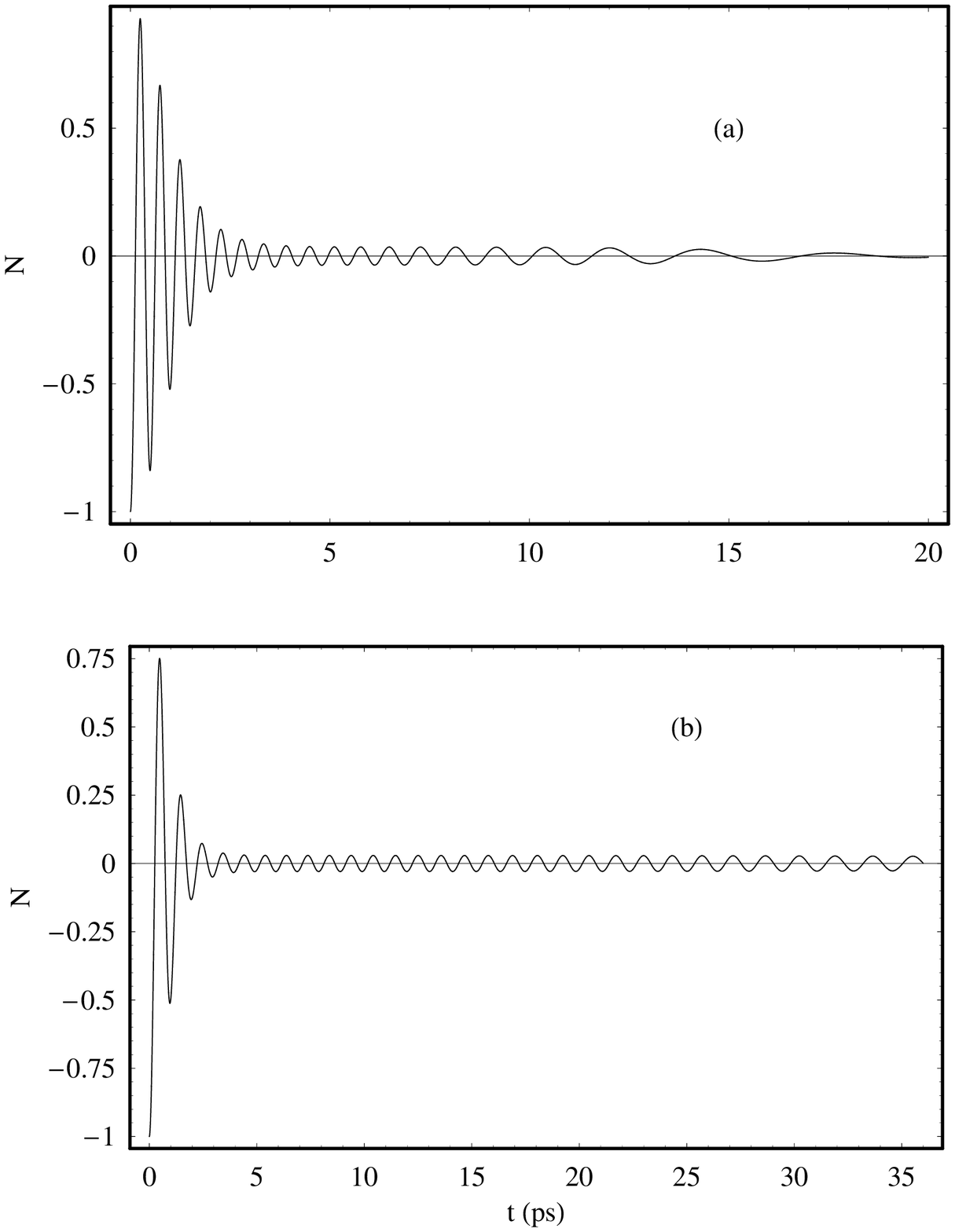}
 \caption{Plots of exciton inversion versus time for two different
 values of pulse duration:  (a) $a=10 ps$, (b) $a=40 ps$.
 Material parameters are pointed out in the text and $T=30 K$.} \label{f1}
\end{center}
\end{figure*}

\begin{figure*}
\begin{center}
\includegraphics[angle=0,width=.5\textwidth]{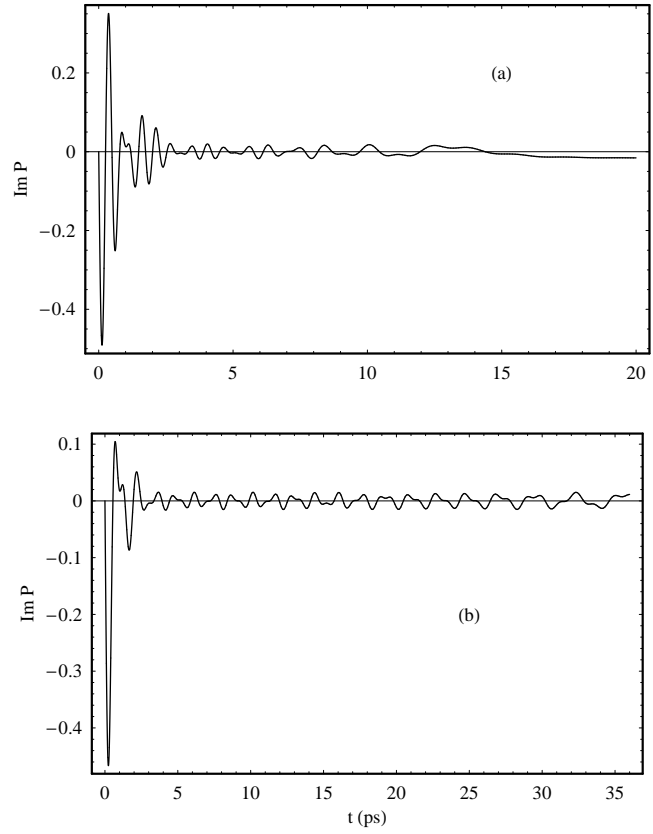}
 \caption{Plots of imaginary part of optical polarization versus time for two different values of
 pulse duration: (a) $a=10 ps$, (b) $a=40 ps$.
 Material parameters are pointed out in the text and $T=30 K$.} \label{f2}
\end{center}
\end{figure*}

In this section we consider the optical coherence of a driven QD
under a pump pulse. Here we neglect the cavity mode and we consider
optical coherence and exciton population dynamics under pulse
excitation and effects of physical parameters such as pulse
duration on these physical quantities. Then the density matrix of
the excitonic system satisfies the following equation of motion
\begin{eqnarray}\label{}
   \dot{\rho}_{ex}(t)&=&\frac{i}{\hbar}[\rho_{ex}(t),\hbar(\hat\sigma_{eg}\alpha(t)+\hat\sigma_{ge}\alpha^\ast(t))]\\ \nonumber
    &-&\int_0^t
   ([\hat\sigma_{ee},\hat\sigma_{ee}\rho(t)]K(t-t')\nonumber \\&-&[\hat\sigma_{ee},\rho(t)\hat\sigma_{ee}]K^\ast(t-t'))dt',
\end{eqnarray}
where $\alpha(t)=f(t)e^{i\omega_{ex}t}$. Exciton population and
optical induced coherence in the QD system are defined through the 
different matrix elements of the density matrix. Exciton
population and optical coherence are defined with the following
set of equations, respectively
\begin{eqnarray}\label{}
    \dot{P}(t)&=&i\alpha(t)(2N_e(t)-1)-P(t)\int_0^tK(t-t')dt',\nonumber\\
    \dot{N}_e(t)&=&2iIm(\alpha^\ast(t)P(t)),
\end{eqnarray}
where $P(t)=\langle e|\hat\rho_{ex}(t)|g\rangle$ and
$N_e(t)=\langle e|\hat\rho_{ex}(t)|e\rangle$. We assume at $t=0$
the QD be in its ground state and at this time it is excited with
a Gaussian pulse excitation with envelop function
$f(t)=\frac{A}{\sqrt{2\pi}a}e^{-\frac{t^2}{a^2}}$ where $a$ is the pulse width and $A$ is a measure of pulse amplitude. For numerical
integration of this set of equations, we shall take a GaAs QD
with a spherical shape. In this case the spectral density is given
by
\begin{equation}\label{}
    j(\omega)=\frac{(\sigma_e-\sigma_h)^2}{4\pi^2\rho c^5}\omega^3
    e^{-\frac{3l^2}{2c^2}\omega^2},
\end{equation}
where $\sigma_e$ and $\sigma_h$ are the bulk deformation potential
constants for electron and hole, $c$ is the sound velocity in the
sample and $l$ is the electron and hole ground-state localization
length  (we assume a spherically symmetric harmonic confinement
potential for the QD and electron and hole in the ground state). We use the following numerical values $\sigma_e-\sigma_h=9
eV$, $\rho=5350 \frac{kg}{m^3}$, $c=5150 \frac{m}{s}$ and $l=4.5
nm$ (these material parameters are approximately acquired from \cite{amk}). Figure (\ref{f1}) shows plots of the time evolution of the  exciton
inversion for two values of pulse duration. In first picoseconds
of dynamics the time evolution shows a strong decrease of exciton
inversion due to the phonon effects and then we see a stable
oscillation in inversion behavior during the pulse duration. It is
clear from the figure that the phonon effects can prevent exciton
generation. On the other hand, we see the complex behavior on the
same timescales of initial dynamics for each pulse duration and
after that small oscillations will continue at the end of pulse
duration. Then we conclude that in the first steps of dynamics the
influence of phonons is a very important damping effect.
Figure(\ref{f2}) shows plots of $Im P(t)$ to consider the time
evolution of optical coherence. As in the case of exciton
population, optical coherence experiences a very rapid decrease
during some first picoseconds. After this strong decrease we see a
very small stable oscillations in optical coherence. Therefore, we
concloud phonon effects are very important on timescales smaller than
the spontaneous decay time and we can consider phonon reservoir as
dominant damping source during the first steps of dynamics.
\section{Interaction of QD with cavity mode}
\begin{figure*}
\begin{center}
\includegraphics[angle=0,width=.5\textwidth]{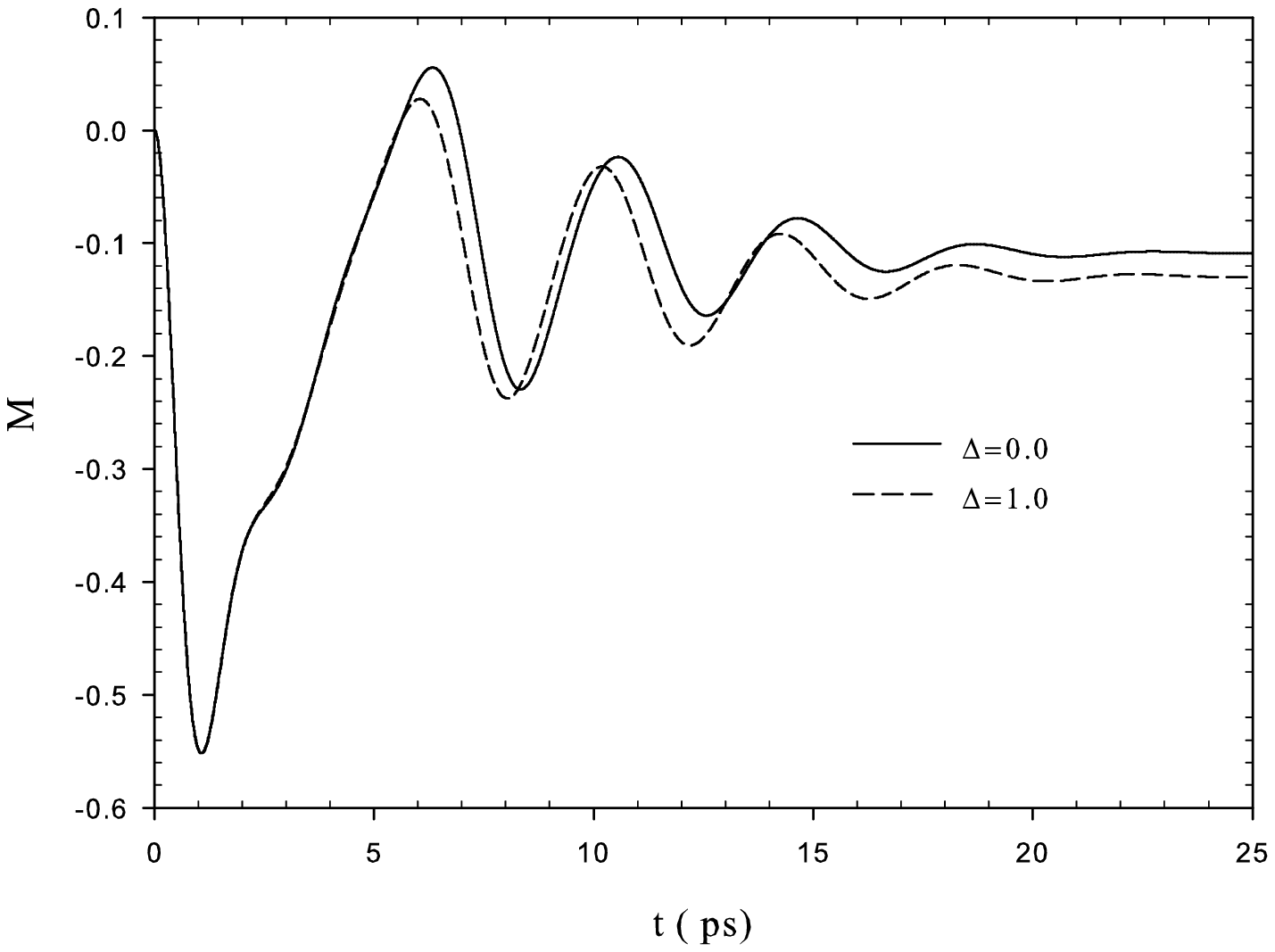}
 \caption{Mandel parameter versus time for pulse duration $a=10 ps$ and $T=30K$ for two different values of
 detuning $\Delta=0.0,\;\Delta=1.0$.} \label{f3}
\end{center}
\end{figure*}

\begin{figure*}
\begin{center}
\includegraphics[angle=0,width=.5\textwidth]{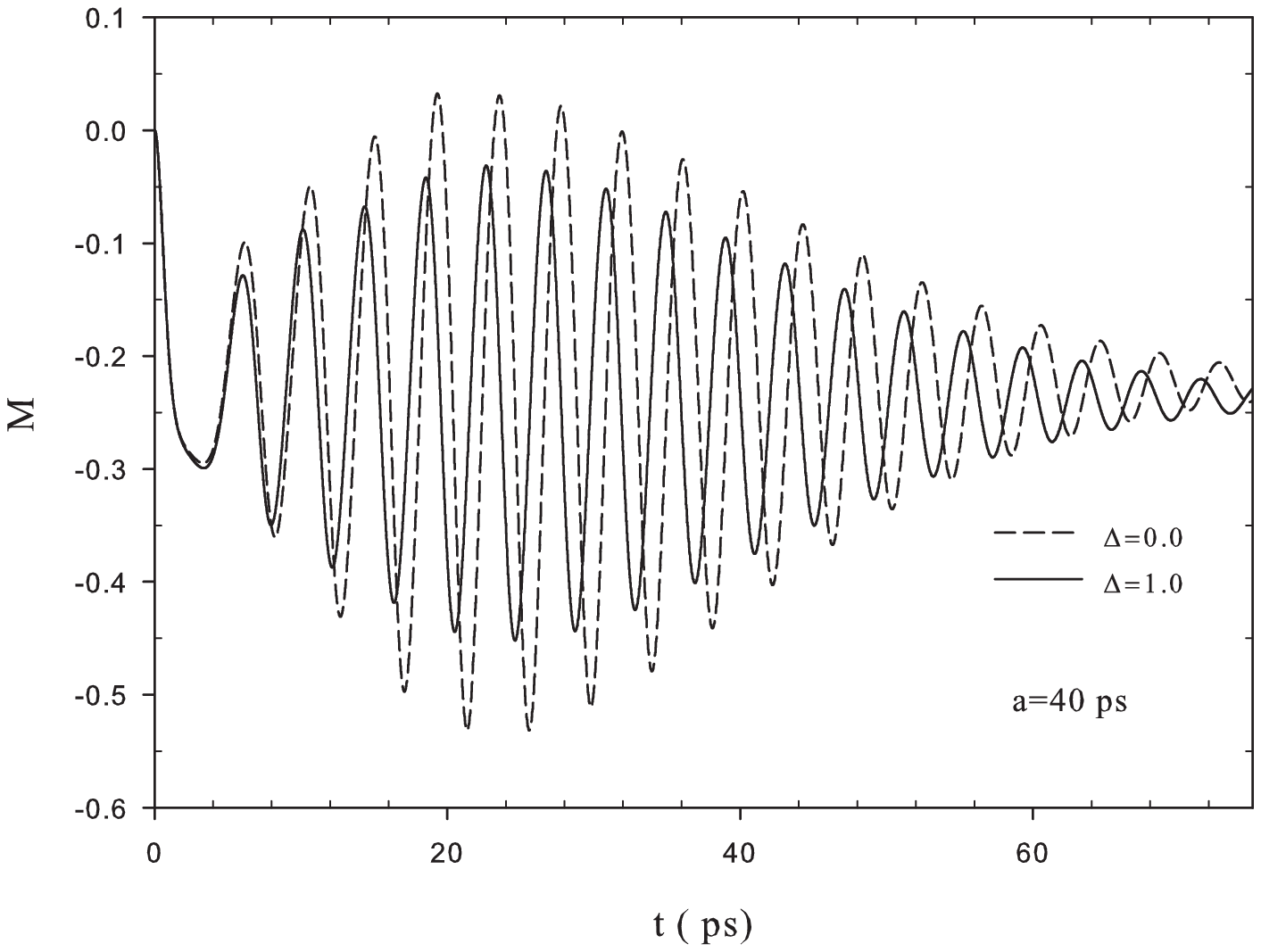}
 \caption{Mandel parameter versus time for pulse duration $a=40 ps$ and $T=30K$ for two different values of
 detuning $\Delta=0.0,\;\Delta=1.0$.} \label{f4}
\end{center}
\end{figure*}

\begin{figure*}
\begin{center}
\includegraphics[angle=0,width=.5\textwidth]{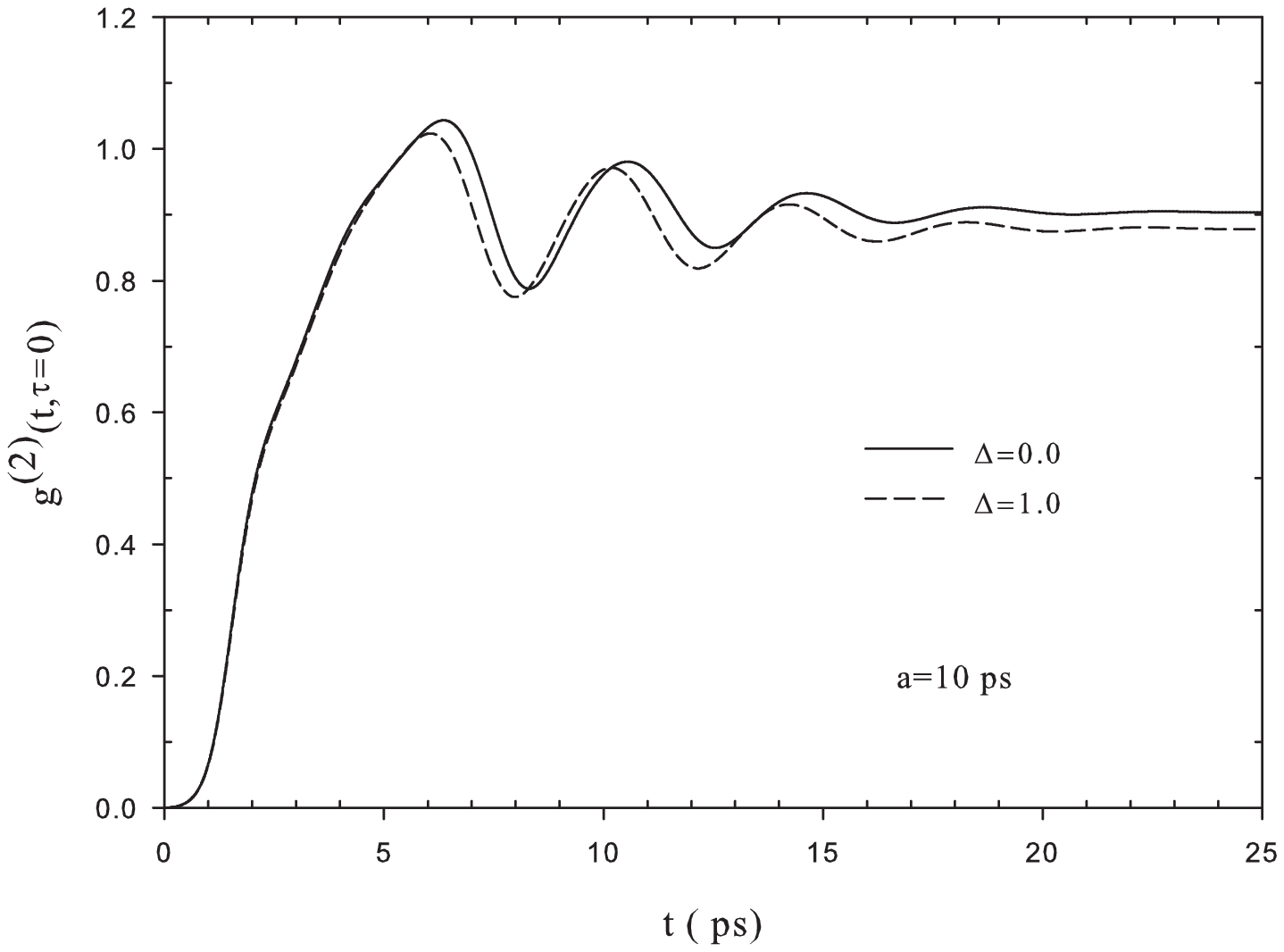}
 \caption{$g^{(2)}(t,\tau=0)$ as a function of time for pulse duration $a=10 ps$ and  $T=30K$ for two
 different values of
 detuning $\Delta=0.0,\;\Delta=1.0$.} \label{f5}
\end{center}
\end{figure*}

\begin{figure*}
\begin{center}
\includegraphics[angle=0,width=.5\textwidth]{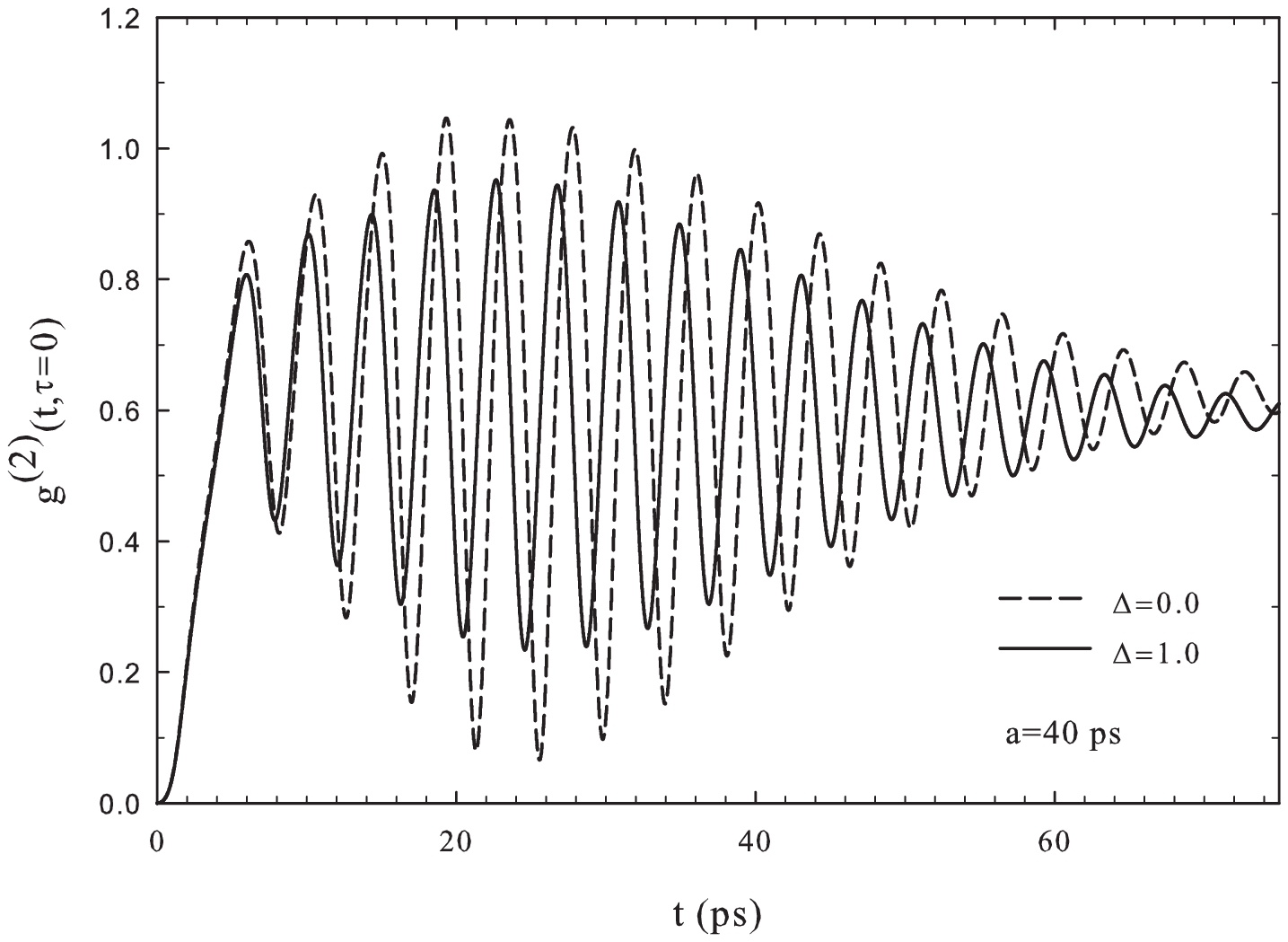}
 \caption{$g^{(2)}(t,\tau=0)$ as a function of time for pulse duration $a=40 ps$ and  $T=30 K$ for two different values
  of detuning $\Delta=0.0,\;\Delta=1.0$.} \label{f6}
\end{center}
\end{figure*}

\begin{figure*}
\begin{center}
\includegraphics[angle=0,width=.5\textwidth]{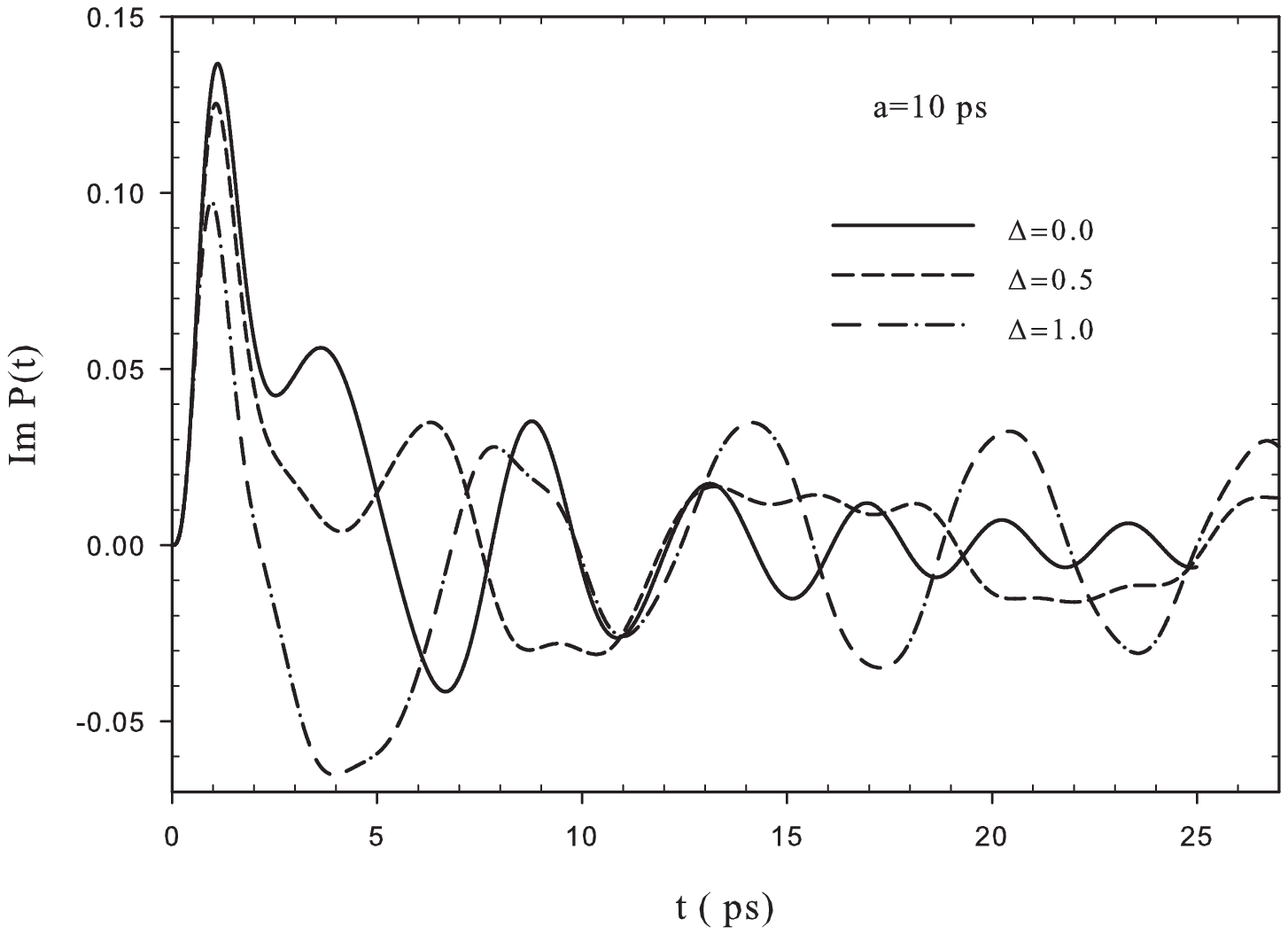}
 \caption{$Im P(t)$ as a function of time for three different values of detuning and for pulse duration $a=10 ps$.
  In this plot $T=30 K$.} \label{f7}
\end{center}
\end{figure*}

\begin{figure*}
\begin{center}
\includegraphics[angle=0,width=.5\textwidth]{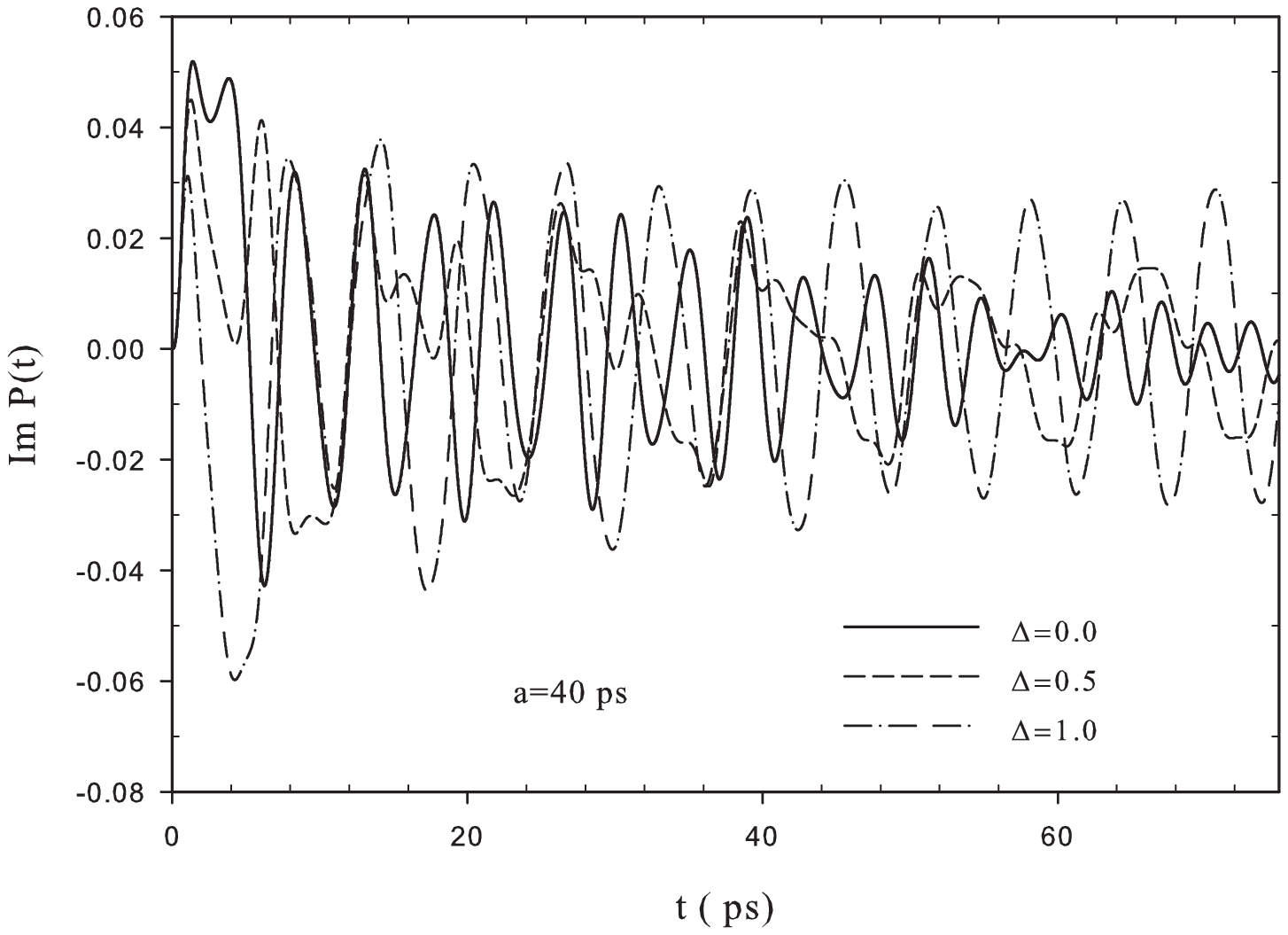}
 \caption{$Im P(t)$ as a function of time for three different values of detuning and for pulse duration $a=40 ps$.
  In this plot $T=30 K$.} \label{f8}
\end{center}
\end{figure*}

In this section we consider the interaction of the QD embedded in
a microcavity with cavity mode. In this case, the density matrix
for the system satisfies Eq.(\ref{eq1}). By using Eq.(\ref{eq1})
one can get a set of differential equations that describe the
evolution of the populations and coherence of the cavity-QD
system. In the basis of product states between the QD states and
Fock states of the cavity mode ($|en\rangle$, $|gn\rangle$) we
calculate the matrix elements of the exciton-photon density
matrix. By taking the matrix elements in Eq.(\ref{eq1}) we get the
following set of linear differential equations for the populations
and coherence in the QD-photon system (we have used the notation $\rho_{in,jm}=\langle in|\rho|jm\rangle$ in which $i$ and $j$ refer to QD states)
\begin{subequations}\label{eq2}
\begin{align}
&\dot{\rho}_{en-1,en-1}(t)=ig\sqrt{n}(
\rho_{en-1,gn}(t)e^{-i\Delta t}-
\rho_{gn,en-1}(t) e^{i\Delta t})\nonumber \\&+if(t)(
\rho_{en-1,gn-1}(t)e^{-i\omega_{ex}t}-
\rho_{gn-1,en-1}(t)e^{i\omega_{ex}t}),\\
&\dot{\rho}_{gn,gn}(t)=ig\sqrt{n}(
\rho_{gn,en-1}(t) e^{i\Delta t}-
\rho_{en-1,gn}(t) e^{-i\Delta t})\nonumber \\&+if(t)(
\rho_{gn,en}(t) e^{i\omega_{ex}t}-
\rho_{en,gn}(t) e^{-i\omega_{ex}t}),\\
&\dot{\rho}_{en-1,gn}(t)=ig\sqrt{n}(
\rho_{en-1,en-1}(t) e^{i\Delta t}\nonumber \\&-
\rho_{gn,gn}(t) e^{i\Delta t})-
\rho_{en-1,gn}(t)\int_0^tK(t-t')dt',\\
&\dot{\rho}_{en-1,gn-1}(t)=if(t)(
\rho_{en-1,en-1}(t) e^{i\omega_{ex}t}\nonumber \\&-
\rho_{gn-1,gn-1}(t) e^{i\omega_{ex}t})-
\rho_{en-1,gn-1}(t)\int_0^tK(t-t')dt'.
\end{align}
\end{subequations}
In the absence of pulse excitation, the matrix elements $
\rho_{en-1,en-1}(t)$, $\rho_{gn,gn}(t)$,
$\rho_{en-1,gn}(t)$ and $
\rho_{gn,en-1}(t)$, for a given photon number, satisfy a
closed set of differential equations. However, the excitation
pulse couples the different terms to each other and an infinite
set of equations has to be solved. In the process of obtaining the
above set of equations we neglect the terms like $
\rho_{gn,gn-1}(t)$ and $\rho_{en,en-1}(t)$
because these terms do not have physical meaning related to the
conditions under consideration. These terms show a coherence in
photon field while the QD remains in its state. This could be
related to photon damping which we have neglected such kind of
terms. On the other hand, we maintain terms like $
\rho_{en,gn}(t)$ which describe coherence in QD system while
photon number is constant. As is clear from (\ref{eq2}) these
terms cam be generated during the dynamics by the pump pulse. \\
\indent As initial condition we take at $t=0$ the QD in its ground
state and cavity field in the vacuum state $
\rho_{g0,g0}(0)=1$, and all other elements of the density
matrix equal to zero. For the numerical integration, the set of
equations can be truncated at a given value, which we take it
equal to 90
 (this value is choose with this condition that the results not change
 with increasing the number of equation).\\
\indent Photon statistics and material characteristics such as
inversion population and optical coherence can be obtained from
(\ref{eq2}). At first we consider Mandel parameter of the cavity
field which is defined as \cite{mandel}
\begin{equation}\label{}
    M=\frac{\langle\hat n^2\rangle-\langle\hat n\rangle^2}{\langle
   \hat n\rangle}-1.
\end{equation}
This parameter vanishes for the Poisson distribution, is positive
for the super-Poisson distribution (photon bunching effect), and
is negative for the sub-Poisson distribution (photon anti-bunching
effect). The mean number of photons in the cavity is (other
moments of $\hat n$ can be calculated in the same manner)
\begin{equation}\label{}
    \langle\hat n\rangle=\sum_n n\left[\rho_{en,en}(t)+\rho_{gn,gn}(t)\right].
\end{equation}
Mandel parameter for the case of resonant interaction ($\Delta=0$)
and in the presence of detuning is plotted, respectively, in
figures (\ref{f3}) and (\ref{f4}) for two different values of pulse
duration. As is seen, the cavity field mode exhibits non-classical
(sub-Poissonian statistics) in the course of time evolution.
Another important feature of this plot is the oscillatory behavior of
Mandel parameter for time scales approximately two times of pulse
duration. Therefore, the emitted photons to cavity mode by QD in
the course of the excitation duration can be reabsorbed by QD and
re-excite the QD then after the end of pulse duration we have oscillations
in photon statistics. On the other hand, it is clear that with
increasing the detuning feature the
amplitude of oscillations in Mandel parameter decrease. \\
\indent Another important quantity in photon statistics is second
order coherence function, $g^{(2)}(t,\tau)$ \cite{qo1},\cite{mandel} which is a
two-time correlation function. Here we consider this quantity for
the case of zero time delay, $g^{(2)}(t,\tau=0)$. This quantity
can be used as an indication of the possible coherence of the
state of the photon system. For the single mode cavity field
$g^{(2)}(t,\tau=0)$ has the following definition
\begin{eqnarray}\label{}
g^{(2)}(t,\tau=0)&=&\frac{\langle a^\dag a^\dag aa\rangle}{\langle a^\dag a\rangle^2}\\ \nonumber&=&\frac{\sum_nn(n-1) \left[\rho_{en,en}(t)+\rho_{gn,gn}(t)\right]}{\left(\sum_nn\left[\rho_{en,en}(t)+\rho_{gn,gn}(t)\right]\right)^2}.
\end{eqnarray}
In the case of resonant interaction and off-resonant interaction,
the plots of this quantity are shown in
figures(\ref{f5}) and (\ref{f6}), respectively. The figures show
non-classical nature of emitted photons (photon anti-bunching).
This quantity shows similar oscillatory behavior to the Mandel
parameter and its oscillatory behavior continue up to times twice
the pulse duration. According to these plots the detuning effects
on $g^{(2)}(t,\tau=0)$ are similar to its effects on the Mandel
parameter and cause the amplitude of oscillation be reduced.
Therefore, in this conditions without any restriction on physical
parameters (damping coefficients and coupling constant) it is
possible that QD emits anti-bunched photons with sub-Poissonian
statistics. The possibility of emitting anti-bunched photons with
sub-Poissonian statistics by a single QD has been considered
experimentally \cite{becher}.\\
\indent The time evolution of the QD coherence in the process of one photon interaction $P(t)=\langle
e0|\rho(t)|g1\rangle$ is shown in figures(\ref{f7}) and (\ref{f8}) for
different values of pulse duration and detuning. In these figures
we plot imaginary part of $P(t)$. These figures indicating
occurrence of decoherence (damping of the imaginary part of
polarization) in the system. The main source of this decoherence
is phonon interaction. In the case of pulse with long duration we
see an irregular oscillation in some time periods. It is clear
that detuning prevents the coherence in this system. However the
detuning is increased the imaginary part of coherence $P(t)$ and
increasing of detuning leads to the regular oscillatory behavior and
causes damping will decrease. In turn, because of the detuning, which weakens the dynamics, the pumping should be increased. Hence these two parameter can be considered as some experimental parameters for controling the decoherence in the QD systems on the timescales
under consideration. On the other hand, by comparison
Fig.(\ref{f2}) with Fig.(\ref{f7}) we can conclude that while the
QD interacts with a cavity mode its optical coherence between
energy levels has a longer life time. Then this can be considered
as another experimental condition for controlling of optical
coherence.
\section{Conclusion}
In this paper we have considered phonon effects (dephasing
effects) on optical properties of a pulse driven QD. We have shown
that these effects strongly affect the Rabi oscillations and
optical coherence. In the time scales which spontaneous emission
and non-radiative recombination do not play an important role in the
dynamics (characteristic times of these effects are much longer
than the characteristic time of phonon reservoir) the phonons
strongly affect optical properties of QD. In the case of the interaction of
system under consideration with cavity mode we have shown that
emitted photons are anti-bunched and obey the sub-Poissonian
statistics. Then in the microcavity with high quality factor which
contains a single QD it is possible to generate non-classical
light in the first some ten picoseconds. Here, we have considered
a Gaussian pulse as a pump. We have shown that with the ending of
pump, oscillations in the photon statistics continue until times
twice the pulse duration. This relates to cavity photon which
remains in the cavity and after ending of pump re-excites the QD.
On the other hand, we have considered the detuning effect on the
optical coherence of QD and we have seen that detuning can prevent
decoherence effects. Hence, detuning can be considered as a
controlling parameter of optical coherence. While QD interacts
resonantly with the cavity mode, we have found that its optical
coherence has a longer life time in comparison with its
interaction with classical pulse. Then by putting the QD in the
cavity it can maintain its coherence between energy levels.
Therefore, the off-resonant interaction of a QD with cavity mode
can be considered as an experimental tool for suppressing
decoherence effects on the exciton.

\textbf{Acknowledgment} The authors wish to thank
      the Office of Graduate Studies of the University of Isfahan and
      Iranian Nanotechnology initiative for
      their support.

{}

\end{document}